\documentclass[11pt]{article}
\usepackage{amssymb}
\usepackage{amsmath}
\usepackage{amstext}
\usepackage{graphicx,epsfig}
\usepackage{epsfig}
\usepackage{verbatim} 
\usepackage{fancybox}
\usepackage{color}
\usepackage{ulem}
\usepackage{enumitem}
\usepackage{subfigure}
\usepackage{bbm}
\usepackage{parskip}
\usepackage{dsfont}
\usepackage[numbers,sort&compress]{natbib}

\newcommand{\Comment}[1]{{}}
\definecolor{darkblue}{rgb}{0.15,0.35,0.55}
\definecolor{reddish}{rgb}{0.65, 0.2, 0.2}
\usepackage[linktocpage=true]{hyperref}
\hypersetup{
colorlinks=true,
citecolor=darkblue,
linkcolor=reddish,
urlcolor=darkblue,
pdfauthor={},
pdftitle={},
pdfsubject={}
}

\newcommand{\be}{\begin{equation}}
\newcommand{\ee}{\end{equation}}
\newcommand{\bea}{\begin{eqnarray}}
\newcommand{\eea}{\end{eqnarray}}
\newcommand{\beas}{\begin{eqnarray*}}
\newcommand{\eeas}{\end{eqnarray*}}
\newcommand{\nn}{\nonumber}

\def\({\left(}
\def\){\right)}

\newcommand{\Tr}{\text{Tr}}

\newcommand{\xb}{\mathbf{x}}

\newcommand{\rd}{{\rm d}}

\def\gsim{ \lower .75ex \hbox{$\sim$} \llap{\raise .27ex \hbox{$>$}} }
\def\lsim{ \lower .75ex \hbox{$\sim$} \llap{\raise .27ex \hbox{$<$}} }

\title{}
\author{}

\numberwithin{equation}{section}

\begin{document}
~
\vspace{2truecm}
\begin{center}
{\LARGE \bf{Goldstones with Extended Shift Symmetries}}
\end{center} 

\vspace{1truecm}
\thispagestyle{empty}
\centerline{{\Large Kurt Hinterbichler${}^{\rm a,}$\footnote{e-mail: \href{mailto:khinterbichler@perimeterinstitute.ca}{khinterbichler@perimeterinstitute.ca}} and Austin Joyce${}^{\rm b,}$\footnote{e-mail: \href{mailto:ajoy@uchicago.edu}{ajoy@uchicago.edu}}}}
\vspace{.7cm}

\centerline{{\it ${}^{\rm a}$Perimeter Institute for Theoretical Physics,}}
 \centerline{{\it 31 Caroline St. N, Waterloo, Ontario, Canada, N2L 2Y5}} 
 \vspace{.5cm}

\centerline{\it ${}^{\rm b}$Enrico Fermi Institute and Kavli Institute for Cosmological Physics,}
\centerline{\it University of Chicago, Chicago, IL 60637}

 \vspace{.8cm}
\begin{abstract}
\noindent
We consider scalar field theories invariant under extended shift symmetries consisting of higher order polynomials in the spacetime coordinates.  These generalize ordinary shift symmetries and the linear shift symmetries of the galileons.  We find Wess--Zumino Lagrangians which transform up to total derivatives under these symmetries, and which possess fewer derivatives per field and lower order equations of motion than the strictly invariant terms.  In the non-relativistic context, where the extended shifts are purely spatial, these theories may describe multi-critical Goldstone bosons.  In the relativistic case, where the shifts involve the full spacetime coordinate, these theories generally propagate extra ghostly degrees of freedom. 
\end{abstract}

\newpage

\tableofcontents
\newpage

\section{Introduction}

The scalar field, $\phi$, with a shift symmetry,
\be \phi\longmapsto \phi+c,\label{shiftsymmetry0}\ee
where $c$ is constant, is the canonical example of a Goldstone boson.  It non-linearly realizes a global internal symmetry.  The invariant Lagrangians are constructed from the first derivative of the field, $\partial\phi$, and possibly additional derivatives, so that each term has a least one derivative per field.  When constructing invariant actions, as opposed to invariant Lagrangians, there is one more possible term: the tadpole term ${\cal L}\sim \phi$.  The tadpole is a Wess--Zumino term.  It is not strictly invariant under \eqref{shiftsymmetry0} but instead changes by a total derivative.  This term has fewer derivatives per field than the others, and has lower order equations of motion; a term with one derivative per field can be expected to have second order equations of motion, whereas the tadpole has zero-th order equations of motion.

The galileon \cite{Nicolis:2008in} is a scalar which generalizes the shift symmetry to include a shift linear in the spacetime coordinates $x^\mu$,
\be \phi\longmapsto \phi+c+b_\mu x^\mu,\label{shiftsymmetry1}\ee
where $b_\mu$ is a constant vector.  These symmetries are no longer internal -- they do not commute with the Poincar\'e transformations -- but rather combine with the Poincar\'e transformations to form a larger algebra, the galileon algebra, which can be thought of as a five dimensional Poincar\'e algebra in which the speed of light in the fifth dimension is taken to infinity \cite{Goon:2012dy}.  The Lagrangians which are invariant under \eqref{shiftsymmetry1} are functions constructed from the second derivatives of the field $\partial\partial\phi$, and possibly additional derivatives, so that each term has a least two derivatives per field.  When constructing invariant actions, there are more possible terms.  These are the galileon terms, and there are $D$ of them in $D$-dimensions.  As with the tadpole in the case of a shift symmetry, they are not invariant under \eqref{shiftsymmetry1} but instead change by a total derivative.  They have fewer derivatives per field than the strictly invariant terms, and their equations of motion are of lower order; a term with two derivatives per field can be expected to have fourth order equations of motion, whereas the galileon terms have at most second order equations of motion.  This ensures that the galileons do not propagate extra degrees of freedom, and for this reason it is said they are ghost-free.

Here, we consider the generalization to shifts which contain higher powers of $x$,
\be \phi\longmapsto \phi+c^{(0)}+c_\mu^{(1)}x^\mu+c_{\mu\nu}^{(2)}x^\mu x^\nu+\cdots+c_{\mu_1\cdots \mu_N}^{(N)}x^{\mu_1}\cdots x^{\mu_N}.\label{shiftsymmetry2}\ee  
The $c_{\mu_1\cdots \mu_N}^{(N)}$ are constant $N$-index tensors, and are totally symmetric (components of any other symmetry type will vanish against the $x$'s).   Clearly, any term with $(N+1)$ or more derivatives per field will be invariant under an $N$-th order symmetry of this type.  Our main goal will be to construct the generalization of the galileon terms for these extended symmetries, those which are invariant only up to a total derivative and have fewer derivatives per field and lower order equations of motion than any strictly invariant term.

Our motivation for studying these theories comes primarily from the non-relativistic setting.  Recently, \cite{Griffin:2013dfa} studied models with ``multi-critical'' non-relativistic Goldstone bosons\footnote{For more on non-relativistic Goldstone bosons, see~\cite{Nielsen:1975hm,Nicolis:2012vf,Watanabe:2012hr,Watanabe:2014fva}.} in which the dispersion relation
\be
\omega^2(\vec k)\sim a_2 \vec k^2+a_4 \vec k^4+\cdots~,
\ee
 has a sound speed squared $\sim a_2$ which can be very small in a technically natural fashion, so that the dispersion relation starts at essentially ${\cal O}(\vec k^4)$.   The small sound speed is due to a naturally small gradient term, $(\nabla\phi)^2$, in the Lagrangian for the Goldstone mode, whose smallness was conjectured to be stabilized by a quadratic higher shift symmetry $\phi\mapsto \phi+c_{ij}x^ix^j$, which forbids the $(\nabla\phi)^2$ gradient term.   The $\vec k^4$ piece is due to a term $(\nabla^2\phi)^2$ which is invariant, up to a total derivative, under the quadratic shift. We will see that this four derivative term is indeed a Wess--Zumino term for the $N=2$ higher shift symmetry -- it has fewer than 3 derivatives per field and transforms by a total derivative.
 Examples of systems with Lorentz non-invariant low energy fixed points which feature a ${\cal O}(\vec k^4)$ dispersion include the ghost condensate~\cite{ArkaniHamed:2003uy}, helical magnets~\cite{PhysRevB.23.4615} and smectic liquid crystals~\cite{PhysRevLett.47.856,PhysRevA26915, cha95}. In these examples, the extended shift symmetry is generally accidental, {\it i.e.}, not respected by the interactions, but there may be cases in which the symmetry persists beyond the free theory.
 
In this paper, we find interacting Wess--Zumino terms.  For example, we will find a cubic term with eight derivatives:
\be
{\cal L}_3\sim \frac{1}{2}\nabla^4\phi (\nabla^2\phi)^2 + \nabla^2\phi (\nabla_i\nabla_j\nabla_k\phi)^2~.
\ee
 In the vicinity of a critical point, such terms will provide the leading extended shift symmetry-respecting irrelevant spatial derivative operators among those with a given number of fields.  They can be expected to determine the leading momentum dependence of certain amplitudes and to enjoy non-renormalization theorems like those of the galileons \cite{Luty:2003vm,Nicolis:2004qq,Hinterbichler:2010xn,deRham:2012ew}.

The general method for constructing actions invariant under non-linearly realized symmetries is the coset construction of Callan, Coleman, Wess and Zumino~\cite{Coleman:1969sm,Callan:1969sn} and Volkov~\cite{Volkov:1973vd}.\footnote{Good general reviews can be found in~\cite{Zumino:1970tu,xthschool}.  For other takes on the construction, including applications, see~\cite{McArthur:2010zm,ArmendarizPicon:2010mz,Hinterbichler:2012mv,Goon:2012dy,Nicolis:2013sga}.}
To construct the Wess--Zumino terms, one must employ a higher-dimensional construction analogous to Witten's construction~\cite{Witten:1983tw} of the Wess--Zumino--Witten term in the chiral Lagrangian~\cite{Wess:1971yu}.  (For more on this construction, including its cohomological interpretation, see~\cite{deAzcarraga:1990gs,D'Hoker:1994ti,deAzcarraga:1995jw,Goon:2012dy}.)
The coset method is applied to the galileons in \cite{Goon:2012dy}.  We review this case and extend it to the more general higher shift symmetries, performing the coset construction and constructing  Wess--Zumino terms. 

All the terms invariant under \eqref{shiftsymmetry2} which we find have equations of motion which are higher than second-order in derivatives.  This is essentially due to the fact that to build terms invariant under the extended shift symmetries we need more than two derivatives per field, and so there is no way for the equations of motion to remain second order.  Therefore, unlike the galileons, if they are considered as Lorentz-invariant theories they will generically propagate ghosts.  (Even so, the relativistic case of the higher shift symmetry finds application in the study of massive higher spin fields \cite{Porrati:2008ha}.)  In the non-relativistic setting this is not an obstruction -- we are interested in terms invariant with respect to a shift of spatial coordinates only, so all the higher derivatives are spatial, and the ghost issue does not arise.   For this reason, we will use non-relativistic notation in the remainder of the paper, though the metric signature and spacetime dimension are left arbitrary.

\noindent
{\bf Conventions:}
In the following, we will be noncommittal about the metric signature, since none of the manipulations we undertake depend on it, so the results can apply to relativistic theories with full spacetime shifts or to non-relativistic theories with only spatial shifts.  Latin letters $i,j,k,\ldots$ denote the coordinate indices, and the space is flat with metric $\delta_{ij}$.  The space (or spacetime in the relativistic case) dimension is $D$.   Our convention for symmetrizing is to use weight one, {\it i.e.}, $(i_1\cdots i_N)={1\over N!}$(sum over permutations of $i$'s).

\section{$N=1$: Galileons}

We start by reviewing the coset construction of the galileons developed in \cite{Goon:2012dy}.  The galileons have a shift symmetry of the form
\be
\phi\longmapsto \phi+c+b_i x^i~.
\label{s1galsymms}
\ee
It is straightforward to construct terms invariant under this symmetry -- indeed any term of the form $(\nabla\nabla\phi)^n$ will do -- but the galileons are distinguished in that they are not strictly invariant under this symmetry.  Rather, the galileon Lagrangians transform up to a total derivative under this shift, leaving the action invariant.  In addition, the galileons are distinguished from the strictly invariant terms in that they have fewer derivatives per field, are not renormalized to any order in perturbation theory \cite{Luty:2003vm,Nicolis:2004qq,Hinterbichler:2010xn,deRham:2012ew}, and have second order equations of motion.  In $D$-dimensions, there are $(D+1)$ galileon terms (including the tadpole).  For example, in $D=4$ they take the form
\bea  
\label{galileonterms}
{\cal L}_1&\sim &\phi\ , \nonumber \\
{\cal L}_2&\sim&(\nabla\phi)^2 \ ,\nonumber \\
{\cal L}_3&\sim&(\nabla {\phi})^2\nabla^2 {\phi} \ ,\nonumber\\
{\cal L}_4&\sim&(\nabla\phi)^2\left[(\nabla^2\phi)^2-(\nabla_i\nabla_j\phi)^2\right] , \nonumber \\
{\cal L}_5&\sim& (\nabla\phi)^2\left[(\nabla^2\phi)^3+2(\nabla_i\nabla_j\phi)^3-3\nabla^2\phi(\nabla_i\nabla_j\phi)^2\right] .\label{gal4terms}
\eea

In order to construct these terms from an algebraic perspective, the first step is to identify the algebra of symmetries.  Infinitesimally, the symmetries act as
\be
\delta_C\phi = 1~,~~~~~~~~~~~~~~~~~~~\delta_{B^i}\phi = x^i~.
\label{galsymms}
\ee
From this, we can deduce the algebra obeyed by these symmetries, which consists of the translations $P^i$ and rotations/boosts $J^{ij}$ of the Poincar\'e algebra plus the generators \eqref{galsymms} of the shift and galileon symmetries.  The non-vanishing commutators are
\begin{align}
\nonumber
&\left [P_{i },B_{j  }\right ] = \delta_{i j }C~,~~~~~~~~~~~~~~~~~~~~~~\left [J_{ij},P_{k }\right ] =\delta_{ik}P_{ j } - \delta_{ jk }P_{i }~,\\ \label{galalgebra}
&\left [J_{ij},B_{k }\right ] =\delta_{ik}B_{ j } - \delta_{ jk }B_{i }~,~~~~~~~~\left [J_{i j },J_{k  l }\right ] = \delta_{i k }J_{j  l }-\delta_{j k }J_{i l }+\delta_{j l }J_{i k }-\delta_{i l }J_{j k }~.
\end{align}
The symmetries corresponding to $B_i$ and $C$ are non-linearly realized on the field, whereas those of the Poincar\'e sub-algebra are linearly realized, so we can think of the galileons as Goldstone fields for the symmetry breaking pattern which takes the full galileon algebra down to the Poincar\'e algebra.

We next introduce fields $\phi(x)$ and $\xi^i(x)$ corresponding to the broken generators, and these parameterize an element of the coset of the symmetry breaking pattern as follows,
\be
g(x) = e^{x^i P_i}e^{\phi(x) C}e^{\xi^i(x) B_i}~.
\ee
From this we construct the Lie algebra-valued Maurer--Cartan 1-form,
\be
\omega =g^{-1}\rd g= \rd x^i P_i+\left(\rd \phi+\xi_i\rd x^i\right)C+\rd\xi^i B_i~.
\label{standardcoset}
\ee
The objects we have at our disposal to build invariant Lagrangians are the 1-forms in the decomposition of the Maurer--Cartan form,
\begin{align}
\omega_P^i = \rd x^i~,~~~~~~~~~~~~\omega_C = \rd \phi+\xi_i\rd x^i~,~~~~~~~~~~~~\omega_B^i = \rd\xi^i~.
\label{Maurer--Cartan1forms}
\end{align}
These 1-forms have nice transformation properties under the galileon symmetries, so it is straightforward to construct invariant actions using them: $\omega_P^i$ provides a vielbein and hence a metric and measure with which to construct actions (just the ordinary flat metric in this case), and the remaining forms provide the basic covariant derivative through which to introduce the Goldstone fields (see~\cite{Goon:2012dy} for more details).  

At this point an additional subtlety arises.  Since we have broken spacetime symmetries, it is possible to eliminate the Goldstone field $\xi^i$ in favor of $\phi$.  In the literature, this often goes by the name {\it inverse Higgs effect}~\cite{Ivanov:1975zq}.\footnote{For various perspectives on the inverse Higgs effect, see~\cite{Volkov:1973vd,Nielsen:1975hm,Low:2001bw,McArthur:2010zm, Hidaka:2012ym, Watanabe:2012hr, Nicolis:2012vf,Nicolis:2013sga, Brauner:2014aha}.} If the commutator of a broken symmetry generator, $Z_1$, with a unbroken translation is proportional to another broken symmetry generator $Z_2$, {\it i.e.} $[P , Z_1] \sim Z_2~$,
then the Goldstone field corresponding to $Z_1$ can be traded for derivatives of the field corresponding to $Z_2$. In practice, this is implemented by setting the 1-form corresponding to $Z_2$ to zero, giving a relation between the Goldstone fields. This is a covariant constraint which allows us to eliminate redundant degrees of freedom. Looking at the galileon algebra~\eqref{galalgebra}, we see that the commutator of $P_i$ with $B_j$ is of precisely this form. We are therefore able to trade $\xi_i$ for $\phi$ by setting $\omega_C = 0$, which yields the relation
\be\label{inversehiggsg}
\xi_i = -\nabla_i\phi~.
\ee
Inserting this relation into~\eqref{Maurer--Cartan1forms}, we see that the only remaining building block is the 1-form
\be
\omega_B^i = -\rd x^j\nabla_j\nabla^i\phi~,
\ee
which manifestly involves two derivatives per $\phi$. This will allow us to construct all invariant Lagrangians with at least two derivatives on each field, but we will miss the galileon terms \eqref{gal4terms}, since they contain fewer than two derivatives per field.

Instead, the galileons correspond to $(D+1)$-forms which we think of as living on the space parametrized by the coordinates and the broken generators, $\{x^i,C,\xi^i\}$, and which are closed under the action of the exterior derivative operator $\rm d$, which acts in a manner determined from the commutation relations \eqref{galalgebra} or from the explicit forms \eqref{Maurer--Cartan1forms},
\begin{align}
\rd\omega_P^i &= 0,\nn\\
\rd\omega_C &= \delta_{ij}\omega_B^i\wedge\omega_P^j ,\nn\\
\rd\omega_B^i &=0.
 \label{dexpressionsg}
\end{align}
The non-trivial closed $(D+1)$-forms are~\cite{Goon:2012dy}
\be
\omega_n^{\rm }\sim \epsilon_{i_1\cdots i_D}\omega_C\wedge\omega_B^{i_1}\wedge\cdots\wedge\omega_B^{i_{n-1}}\wedge\omega_P^{i_n}\wedge\cdots\wedge\omega_P^{i_D}~.
\label{gald1form}
\ee
They are invariant under the non-linear symmetries since they are constructed solely from the Maurer--Cartan form \eqref{Maurer--Cartan1forms}.
We can express these as the derivative of a non-invariant $D$-form,\footnote{Note that the closed $(D+1)$-forms are always exact because de Rham cohomology on the spaces under consideration is trivial. 
}
\be
\omega_n^{\rm } = \rd \beta^{\rm }_n~,
\ee
where
\begin{align} 
\beta_n^{\rm } \sim  \epsilon_{i_1\cdots i_D}\bigg(\phi\,\rd \xi^{i_1}\wedge\cdots\wedge\rd \xi^{i_{n-1}}&\wedge\rd x^{i_n}\wedge\cdots\wedge\rd x^{i_D}
\\\nonumber
&-{(n-1)\over 2(D-n+2)}\xi^2 \rd \xi^{i_1}\wedge\cdots\wedge\rd \xi^{i_{n-2}}\wedge\rd x^{i_{n-1}}\wedge\cdots\wedge\rd x^{i_D}\bigg).
\end{align}
We then pull the $D$-form $\beta^{\rm }_n$ back to the physical space via the map $\{x^i, \phi(x), \xi^i(x)\}$ and integrate over $x^i$ to create an action for $\phi,\xi^i$.
\be
S_n \sim \int_{}\,\beta^{\rm }_n \sim \int \rd^Dx\ {\cal L}_n(\phi,\xi)~.
\ee 
Using the inverse Higgs relation \eqref{inversehiggsg} we eliminate $\xi^i$ and obtain a Lagrangian solely for $\phi$,
which reads (after some integrations by parts),
\be
{\cal L}_n \sim \epsilon_{i_1\cdots i_{n-1} k_n\cdots k_D}\epsilon^{j_1\cdots j_{n-1}k_n \cdots k_D} \phi \nabla_{j_1}\nabla^{i_1}\phi\cdots\nabla_{j_{n-1}}\nabla^{i_{n-1}}\phi~.
\label{wzlags1}
\ee
These are the galileon Lagrangians in any dimension.  For $D=4$ they reproduce the expressions \eqref{gal4terms}.

\section{$N=2$: Quadratic Shifts}
We now extend this construction to a polynomial shift symmetry of the field of the form
\be\label{quadtransexp}
\phi\longmapsto \phi+ c+b_i x^i+S_{ij}x^i x^j~,
\ee
where $S_{ij}$ is a constant symmetric matrix.  The is the case $N=2$ of \eqref{shiftsymmetry2}, relevant to the examples in \cite{Griffin:2013dfa}.

We first work out the algebra obeyed by these symmetries.
Infinitesimally, the symmetries act on the field as
\be
\delta_C\phi = 1~,~~~~~~~~~~~~~~~~~~~\delta_{B^i}\phi = x^i~,~~~~~~~~~~~~~~~~~~~\delta_{S^{ij}}\phi = x^i x^j.
\label{x2shiftsymms}
\ee
These are the non-linearly realized symmetries.  The non-vanishing commutators among these generators and the linearly realized Poincar\'e generators are
\begin{align}
\nonumber
\label{x2algebra}
\left [P_i,B_j\right ] &= \delta_{ij }C~,  ~~~~~~~~~~~~~~~~~~~~~~~~~~~~~~ \left[P_{i },S_{jk}\right] = \delta_{ij}B_k+\delta_{ik}B_j~, \\
\left[J_{ij},B_{k }\right]& =\delta_{ik }B_{j }-\delta_{jk}B_{i}~,~~~~~~~~~~~~~~~~~~ \left[J_{ij},S_{kl},\right] =\delta_{ ik}S_{jl}- \delta_{ jk }S_{ il }+\delta_{  il}S_{ k j}-\delta_{  jl}S_{ k i} ~,\\\nonumber
\left [J_{ij },P_{k }\right ] &=\delta_{ik}P_{j } - \delta_{jk }P_{i }~,~~~~~~~~~~~~~~~~~~~\left [J_{ij },J_{kl }\right ] = \delta_{ik}J_{jl }-\delta_{jk }J_{il }+\delta_{jl}J_{ik }-\delta_{il}J_{jk }~.
\end{align}
The first line tells us that $P_i$ acts as a kind of lowering operator for the degree of the shift symmetries.  It also tells us that we cannot have only the higher shifts without the lower shifts, since the lower shifts appear as commutators of the higher shifts with the momentum.  The second line tells us that the shift symmetries transform in the expected way as tensors under rotations, and the final line is the standard Poincar\'e algebra.

\subsection{Coset Construction of Invariant Lagrangians\label{N2secsta}}
Taking the algebra~\eqref{x2algebra}, we want to construct a theory non-linearly realizing the symmetries associated to the broken generators,
\be
C~,~~~~~~~~~~~~~B_i~,~~~~~~~~~~~~~S_{ij}~.
\ee
We introduce fields $\phi(x)$, $\xi^i(x)$ and symmetric $\Phi^{ij}(x)$, and parameterize the coset space as
\be
g(x) = e^{x^i P_i}e^{\phi(x) C}e^{\xi^i (x)B_i}e^{\Phi^{ij}(x)S_{ij}}~.
\ee
From this, we calculate the Maurer--Cartan form,
\be
\omega = g^{-1}\rd g= \rd x^i P_i+(\rd\phi+\xi_i\rd x^i)C+(\rd\xi^i+2\rd x^j \Phi_j^i)B_i+\rd \Phi^{ij}S_{ij}~.
\label{x2Maurer--Cartanforms}
\ee
The building blocks that we may use to build invariant Lagrangians are the 1-forms
\be
\label{x2Maurer--Cartan1forms}
\omega_P^i = \rd x^i~,~~~~~~~~~\omega_C = \rd\phi+\xi_i\rd x^i~,~~~~~~~~~\omega_B^i = \rd\xi^i+2\rd x^j \Phi_j^i~,~~~~~~~~~\omega_S^{ij} = \rd \Phi^{ij}.
\ee

Now, we note that there are two inverse Higgs constraints, coming from the commutators
\be
\left [P_{i },B_{j  }\right ] = \delta_{ij }C~,~~~~~~~~~~~~~\left[P_{i },S_{jk}\right] = \delta_{ij}B_k+\delta_{ik}B_j~,
\ee
which tells us that we can eliminate both $\xi_i$ and $\Phi_{ij}$ by setting both $\omega_B$ and $\omega_C$ equal to zero.\footnote{Strictly speaking, the inverse Higgs constraint tells us that we should set the symmetric part of $\omega_B$ to zero, but in this case $\omega_B$ is symmetric, so the difference here is irrelevant.} From setting $\omega_B = 0$, we obtain the relation
\be\label{invhiggsn21}
\Phi_{ij} = -\frac{1}{2}\nabla_{(i}\xi_{j)}~,
\ee
then, from setting $\omega_C=0$, we have the relation
\be\label{invhiggsn22}
\xi_i = -\nabla_i\phi~.
\ee
Combining these two relations, we find that
\be\label{invhiggsn23}
\Phi_{ij} = \frac{1}{2}\nabla_i\nabla_j\phi~.
\ee
The remaining 1-form we may use to build Lagrangians is
\be
\omega_S^{ij} = \frac{1}{2}\rd x^k\nabla_k\nabla^i\nabla^j\phi~,
\ee
which involves three derivatives on the field.   This tells us that to build Lagrangians {strictly} invariant under the extended shift symmetry \eqref{quadtransexp}, we write any term involving at least three derivatives per field. 

\subsection{Construction of Wess--Zumino Terms}
We are more interested in the terms which are not strictly invariant, but which shift by a total derivative under the symmetry.  These are the Wess--Zumino terms for the symmetries~\eqref{x2shiftsymms}, the analogs of the galileons.
These correspond to $(D+1)$-forms constructed from the building blocks~\eqref{x2Maurer--Cartan1forms} in a rotationally (Lorentz)-invariant way which are annihilated by the exterior derivative, but which { cannot} be written as the exterior derivative of something built out of~\eqref{x2Maurer--Cartan1forms}.

We need to know how the exterior derivative acts on the basis forms.  This is determined from the commutation relations or is readily computed from \eqref{x2Maurer--Cartan1forms},
\begin{align}
\rd\omega_P^i &= 0,\nn\\
\rd\omega_C &= \delta_{ij}\omega_B^i\wedge\omega_P^j ,\nn\\
\rd\omega_B^i &= 2\delta_{j k}\omega_S^{i j}\wedge\omega_P^ k,\nn\\
\rd\omega_S^{ij} &= 0~. \label{dexpressions}
\end{align}
We then look for $(D+1)$-forms on the space with coordinates $\{x^i,\phi,\xi^i,\Phi^{ij}\}$ which are closed, $\rd\omega = 0$.  Given such a form, we write it as the exterior derivative of a (non-invariant) $D$-form: $\omega = \rd\beta$, pull back $\beta$ to the physical $D$-dimensional space given by $\{x^i, \phi(x),\xi^i(x),\Phi^{ij}(x)\}$ and integrate over $x^i$, giving a Wess--Zumino Lagrangian,
\be
S\sim \int_{}\,\beta \sim  \int\rd^Dx \,{\cal L}~.\label{patternint}
\ee
In what follows, we will find closed $(D+1)$-forms and their corresponding Lagrangians.

\subsection{Tadpole Terms}

First is a closed form involving $\omega_C$,
\begin{align}
\omega_1 &=\epsilon_{i_1\cdots i_D}\omega_C\wedge\omega_P^{i_1}\wedge\cdots\wedge\omega_P^{i_D}.
\end{align}
Taking the exterior derivative and using \eqref{dexpressions}, we find an expression with more $\omega_P$'s than can be anti-symmetrized, so the result vanishes.

Using the expressions \eqref{x2Maurer--Cartan1forms}, we write this as an exact form,
\be
\omega_1 = \epsilon_{i_1\cdots i_D}\rd\phi\wedge\rd x^{i_1}\wedge\cdots\wedge\rd x^{i_D} = \rd\beta_1,\ \ \ \ \ \ \beta_1\sim\epsilon_{i_1\cdots i_D}\phi~ \rd x^{i_1}\wedge\cdots\wedge\rd x^{i_D} ~.
\ee
Integrating as in \eqref{patternint}, we obtain the tadpole Lagrangian
\be
{\cal L}_1 \sim \phi~,
\ee
which indeed is invariant up to a total derivative under the symmetries~\eqref{x2shiftsymms}. 

\subsection{Quadratic Terms\label{N2quadtermssec}}

Next is a form involving $\omega_B$,
\begin{align}
\omega_2 &=\epsilon_{i_1\cdots i_D}\left[\omega_S\right]\wedge\omega_B^{i_1}\wedge\omega_P^{i_2}\wedge\cdots\wedge\omega_P^{i_D},~\label{formb}
\end{align}
where brackets $\left[\ldots \right]$ indicate the trace $\delta_{ij}\omega_S^{ij}$.  We can see that this is closed as follows: from \eqref{dexpressions}, we see that when acting with $\rd$, $\omega_B$ will be replaced by an $\omega_P$ and an $\omega_S$.  There will then be $D$ factors of $\omega_P$ which must therefore be proportional to $\epsilon^{i_1\cdots i_D}$.  This new epsilon contracts with the epsilon in \eqref{formb}, creating deltas which then yield an expression with two factors of $\left[\omega_S\right]$, which vanishes by anti-symmetry of the wedge product.

The expression \eqref{formb} can be written as
\be
\omega_2 \sim \epsilon_{i_1\cdots i_D}\rd[\Phi]\wedge(\rd\xi^{i_1}+2\Phi_j^{i_1}\rd x^j)\wedge\rd x^{i_2}\wedge \cdots\wedge \rd x^{i_D}~,
\ee
which is exact $\omega_2 \sim \rd\beta_2$, with\footnote{To obtain this result, it is useful to note that
\begin{equation*}
2\epsilon_{i_1\cdots i_D}\Phi_\ell^{i_1}\rd[\Phi]\wedge\rd x^\ell\wedge\rd x^{i_2}\wedge\cdots\wedge\rd x^{i_D} = 2(D-1)! [\Phi]\rd[\Phi] \rd^Dx = (D-1)! \rd \left([\Phi]^2 \right)\rd^Dx = \frac{1}{D!}
\epsilon_{i_1\cdots i_D}[\Phi]^2\rd x^{i_1}\wedge\cdots\wedge\rd x^{i_D}~.
\end{equation*}
}
\be
\beta_2 \sim \epsilon_{i_1\cdots i_D}\xi^{i_1}\rd[\Phi]\wedge\rd x^{i_2}\wedge\cdots\wedge\rd x^{i_D}+\frac{1}{D!}
\epsilon_{i_1\cdots i_D}[\Phi]^2\rd x^{i_1}\wedge\cdots\wedge\rd x^{i_D}~.
\ee
Integrating, this gives us
\be
S_2 \sim \int \beta_2 \sim \int\rd^4 x\left(-\xi^i\nabla_i[\Phi]+[\Phi]^2\right)~.
\ee
Using the inverse Higgs constraints \eqref{invhiggsn22}, \eqref{invhiggsn23} and integrating by parts
we find the invariant Lagrangian
\be
{\cal L}_2\sim (\nabla^2\phi)^2~.
\ee
It is straightforward to verify that this term is invariant under the quadratic shift symmetries \eqref{quadtransexp}, up to a total derivative.  This is the term that yields an ${\cal O}(\vec k^4)$ contribution to the dispersion relation in the context of \cite{Griffin:2013dfa}.

\subsection{Cubic Terms}

Next we consider the form
\be
\omega_3 \sim\epsilon_{i_1\cdots i_D}\left[\omega_S\right]\wedge\omega_S^{i_1 j}\wedge{\omega_S}^{i_2}_{\ \ j}\wedge\omega_P^{i_3}\wedge\cdots\wedge\omega_P^{i_D}.
\ee
This is closed since it is constructed only from $\omega_P$ and $\omega_S$, which both vanish under $\rd$.  We write it as an exact form,
\be
\omega_3\sim\epsilon_{i_1\cdots i_D} \rd[\Phi]\wedge\rd \Phi^{i_1 j}\wedge\rd \Phi^{i_2}_{\ \ j}\wedge\rd x^{i_3}\wedge\cdots\wedge\rd x^{i_D} = \rd\beta_3~,
\ee
with
\be
 \beta_3\sim\epsilon_{i_1\cdots i_D} [\Phi]\rd \Phi^{i_1 j}\wedge\rd \Phi^{i_2}_{\ \ j}\wedge\rd x^{i_3}\wedge\cdots\wedge \rd x^{i_D}~.
\ee
Integrating,
\be
 S_3 \sim \int \beta_3^{\rm } \sim  \int\rd^D x \,\epsilon_{i_1i_2 k_3\cdots k_D} \epsilon^{j_1j_2k_3\cdots k_D} [\Phi]\nabla_{j_1} \Phi_{ l}^{i_1}\nabla_{j_2} \Phi^{i_2 l}~.
\ee
Imposing the inverse Higgs constraints \eqref{invhiggsn22}, \eqref{invhiggsn23} we arrive at a cubic Lagrangian with 8 derivatives,
\be\label{cubiclag1}
{\cal L}_3 \sim \epsilon_{i_1i_2  k_3\cdots k_D} \epsilon^{j_1j_2 k_3\cdots k_D} \nabla^2\phi \nabla_{j_1}\nabla^{i_1} \nabla_{ l}\phi \nabla_{j_2} \nabla^{i_2}\nabla^{ l}\phi. 
\ee
We can check directly that this is invariant up to a total derivative under the quadratic shift symmetry \eqref{quadtransexp}: only the $\nabla^2\phi$ piece transforms, and the change is a total derivative due to the anti-symmetry of the epsilon structure, $\delta{\cal L}_3\sim \nabla_{j_1}\left(s^m_{\ m} \epsilon_{i_1i_2  k_3\cdots k_D} \epsilon^{j_1j_2 k_3\cdots k_D}  \nabla^{i_1} \nabla_{ l}\phi \nabla_{j_2} \nabla^{i_2}\nabla^{ l}\phi\right)$.

Contracting the epsilons and integrating by parts, we may write \eqref{cubiclag1} as
\be\label{cubicexpandedt}
{\cal L}_3 \sim \frac{1}{2}\nabla^4\phi (\nabla^2\phi)^2 + \nabla^2 \phi (\nabla_i\nabla_j\nabla_ k\phi)^2~.
\ee
Like the galileon, the equations of motion stemming from this Lagrangian are lower order than na\"ively expected.  Since the Lagrangian involves third derivatives, the equations of motion are na\"ively expected to be sixth order.   We can see from the anti-symmetry of the $\epsilon$-symbol that the equations of motion will not contain any sixth derivatives.  What is perhaps less obvious is that the five derivative terms also cancel, and we are left with fourth order equations:
\be {\delta {\cal L}_3\over \delta \phi}\sim (\nabla_i\nabla_j\nabla_ k\nabla_ l\phi)^2-2(\nabla^2 \nabla_i\nabla_j\phi )^2+(\nabla^4\phi)^2.\ee

\subsection{Higher Terms}

As we go up in dimension, there are more possible terms.
There is a quintic term that we can write in $D\geq 4$ that we can not in $D\leq 3$, which stems from the form
\be
\omega_5^{\rm } \sim\epsilon_{i_1i_2i_3i_4i_5\cdots i_D}\left[\omega_S\right]\wedge\omega_S^{i_1j}\wedge{\omega_S}^{i_2}_{\ \ j}\wedge\omega_S^{i_3 k}\wedge{\omega_S}^{i_4}_{\ \  k}\wedge \omega_P^{i_5}\wedge\cdots\wedge \omega_P^{i_D}~.
\ee
This is
\be \omega^{\rm }_5\sim\epsilon_{i_1i_2i_3i_4i_5\cdots i_D}\rd [\Phi]\wedge\rd \Phi^{i_1j}\wedge\rd \Phi^{i_2}_{\ \ j}\wedge\rd \Phi^{i_3 k}\wedge\rd \Phi^{i_4}_{\ \  k}\wedge \rd x^{i_5}\wedge\cdots\wedge \rd x^{i_D}=\rd\beta^{\rm }_5
\ee
with
\be
\beta^{\rm }_5\sim\epsilon_{i_1i_2i_3i_4i_5\cdots i_D}[\Phi]\rd \Phi^{i_1j}\wedge\rd \Phi^{i_2}_{\ \ j}\wedge\rd \Phi^{i_3 k}\wedge\rd \Phi^{i_4}_{\ \  k}\wedge \rd x^{i_5}\wedge\cdots\wedge \rd x^{i_D},
\ee
which leads to a Lagrangian with 5 fields and 14 derivatives,
\be {\cal L}_5\sim \epsilon_{i_1i_2i_3i_4 k_5\cdots k_D}\epsilon^{j_1j_2j_3j_4 k_5\cdots k_D} \nabla^2\phi\nabla^{i_1}\nabla_{j_1}\nabla^{ l}\phi\nabla^{i_2}\nabla_{j_2}\nabla_{ l}\phi\nabla^{i_3}\nabla_{j_3}\nabla^{m}\phi\nabla^{i_4}\nabla_{j_4}\nabla_{m}\phi .
\ee
Expanding out the epsilons and integrating by parts,
\begin{align}
\nonumber
{\cal L}_5 &\sim \nabla^2\phi\nabla_i\nabla^2\phi\nabla_j\nabla^2\phi\nabla^i\nabla_ k \nabla_\ell\phi\nabla^j\nabla^ k \nabla^\ell\phi-\frac{1}{4}\nabla^2\phi\nabla_i\nabla^2\phi\nabla^i\nabla^2\phi\nabla_j\nabla^2\phi\nabla^j\nabla^2\phi\\\nonumber
&+\frac{1}{2}\nabla^2\phi\nabla^i\nabla^j\nabla^k\phi\nabla_i\nabla_\ell\nabla_m\phi\nabla_j\nabla^\ell\nabla^n\phi\nabla_k\nabla_m\nabla_n\phi+\frac{1}{2}\nabla^2\phi\nabla^i\nabla^j\nabla^k\phi\nabla^\ell\nabla_j\nabla_k\phi\nabla_i\nabla_m\nabla_n\phi\nabla^\ell\nabla^m\nabla^n\phi\\\nonumber
&-2\nabla^2\phi\nabla^i\nabla^2\phi\nabla^j\nabla_i\nabla_k\phi\nabla_j\nabla_m\nabla_n\phi\nabla^k\nabla^m\nabla^n\phi+\frac{1}{2}\nabla^2\phi\nabla^i\nabla^2\phi\nabla_i\nabla^2\phi (\nabla_k\nabla_\ell\nabla_m\phi)^2\\
&-\frac{1}{4}\nabla^2\phi(\nabla_i\nabla_j\nabla_k\phi)^2(\nabla_\ell\nabla_m\nabla_n\phi)^2~.
\end{align}
As with the cubic term \eqref{cubicexpandedt}, the equations of motion stemming from this term are fourth order in derivatives.

This pattern of terms continues into higher dimensions: at each jump of 2 in dimension we obtain a new non-trivial term given by adding in another pair of contracted $\omega_S$'s, leading to the $D$-form,
\be
\omega_{n}^{\rm } \sim\epsilon_{i_1\cdots i_D}\left[\omega_S\right]\wedge\omega_S^{i_1j}\wedge{\omega_S}^{i_2}_{\ \ j}\wedge\cdots\wedge\omega_S^{i_{n-2} k}\wedge{\omega_S}^{i_{n-1}}_{\ \  k}\wedge \omega_P^{i_{n}}\wedge\cdots\wedge \omega_P^{i_D}~.
\ee
This yields a term with $n$ powers of $\phi$ in dimension $D\geq n-1$ for $D$ even and $D\geq n$ for $D$ odd.
\be {\cal L}_{n}\sim \epsilon_{i_1\cdots i_{n-1} k_{n}\cdots k_D}\epsilon^{j_1\cdots j_{n-1} k_{n}\cdots k_D} \nabla^2\phi\nabla^{i_1}\nabla_{j_1}\nabla^{ l}\phi\nabla^{i_2}\nabla_{j_2}\nabla_{ l}\phi\cdots \nabla^{i_{n-2}}\nabla_{j_{n-2}}\nabla^{m}\phi\nabla^{i_{n-1}}\nabla_{j_{n-1}}\nabla_{m}\phi .
\ee
Again, the equations of motion are fourth order in derivatives, and we can see directly how these terms are invariant up to a total derivative under the quadratic shift symmetry \eqref{quadtransexp} by using the anti-symmetry of the epsilon.

There is also a parity violating Wess--Zumino $(D+1)$-form,
\be
\omega_p \sim \Tr\left[\omega_S\wedge\omega_S\wedge\cdots\wedge\omega_S\right]\sim \Tr\left[\rd \Phi\wedge \rd \Phi\wedge\cdots\wedge \rd \Phi\right],
\ee
where we have written $\omega_S$ and $\Phi$ with indices suppressed, {\it i.e.}, as symmetric matrix valued forms.  This can be written as $\rd\beta_p$ with
\be
\beta_p \sim \Tr\left[\Phi~\rd \Phi\wedge\cdots\wedge \rd \Phi\right].
\ee
Upon pulling this back to the physical space and substituting the inverse Higgs relation, this becomes the Lagrangian
\be
{\cal L}_p \sim  \epsilon^{i_1\cdots i_D} \Tr\left[\Pi \nabla_{i_1}\Pi \nabla_{i_2}\Pi\cdots \nabla_{i_D}\Pi\right],
\label{3dtotaldWZterm}
\ee
where we have defined the matrix of second derivatives $\Pi_{ij}\equiv \nabla_i\nabla_j\pi$.

Under the quadratic shift symmetry \eqref{quadtransexp}, only the two derivative term changes, and the result is a total derivative,
\be \delta {\cal L}_p\sim   \epsilon^{i_1\cdots i_D} \Tr\left[s \nabla_{i_1}\Pi \nabla_{i_2}\Pi\cdots \nabla_{i_D}\Pi\right]=\nabla_{i_1}\left(\epsilon^{i_1\cdots i_D} \Tr\left[s \Pi \nabla_{i_2}\Pi\cdots \nabla_{i_D}\Pi\right]\right).\ee

For $D=2,3$ \eqref{3dtotaldWZterm} vanishes identically: $\Tr\left[\Pi \nabla_{i_1}\Pi \nabla_{i_2}\Pi\right]$ is symmetric in $i_2,i_2$, as can be seen by reversing and cycling the trace, and $\Tr\left[\Pi \nabla_{i_1}\Pi \nabla_{i_2}\Pi\nabla_{i_3}\Pi\right]$ is symmetric in $i_1,i_3$ for the same reason, so these vanish against the anti-symmetric $\epsilon$-symbol.  For $D\geq 4$ on the other hand, we obtain non-trivial terms.   As with the other terms, the equations of motion contain at most fourth derivatives.

\section{Arbitrary $N$}
We now move on to the case of general $N$, and consider shift symmetries of the form
\be \phi\longmapsto \phi+c^{(0)}+c_i^{(1)}x^i+c_{ij}^{(2)}x^i x^j+\cdots+c_{i_1\cdots i_N}^{(N)}x^{i1}\cdots x^{i_N},\label{shiftsymmetryN}\ee  
where the $c_{i_1\cdots i_N}^{(K)}$, $K=0,\cdots,N$ are constant, symmetric tensors.
The infinitesimal action of these symmetries on the field is
\be
\delta_S\phi=1~,~~~~\delta_{S^i}\phi=x^i~,~~~~\delta_{S^{ij}}\phi=x^i x^j,~~\cdots,~~\delta_{S^{i_1\cdots i_N}}\phi=x^{i_1}\cdots x^{i_N}~.
\ee
The commutation relations among these generators and the Poincar\'e generators are
\begin{align}
[P_j, S_{i_1\cdots i_K}] &= \sum_{\ell=1}^K\delta_{j i_\ell}S_{i_1\cdots \hat i_\ell\cdots i_K},\ \ \ ~~~~~~~K=0,\cdots,N \nn\\
[J_{jk}, S_{i_1\cdots i_K}] &= \sum_{\ell=1}^K\left(\delta_{j i_\ell}S_{k i_1\cdots \hat i_\ell\cdots i_K} - \delta_{k i_\ell}S_{j i_1\cdots \hat i_\ell\cdots i_N}\right),\ \ \ ~~~~K=0,\cdots,N \nn\\
\left [J_{ij },P_{k }\right ] &=\delta_{ik}P_{j } - \delta_{jk }P_{i },~~~~~~~~~~~~\left [J_{ij },J_{kl }\right ] = \delta_{ik}J_{jl }-\delta_{jk }J_{il }+\delta_{jl}J_{ik }-\delta_{il}J_{jk }~,
\end{align}
where the hat means that we omit the corresponding index. 
Generalizing the lower $N$ cases, the first line tells us that the momentum $P_i$ acts as a kind of lowering operator for the degree of the shift symmetries, and also tells us that we cannot have only the higher shifts without the lower shifts, since the lower shifts appear as commutators of the higher shifts with the momentum.  The second line tells us that the shift symmetries transform as tensors under rotations, and the final line is the standard Poincar\'e algebra.

The broken symmetries are $S_{i_1\cdots i_K}$ for $K=0,\cdots,N$.  We introduce symmetric fields $\Phi^{i_1\cdots i_K}(x)$ for $K=0,\cdots,N$ 
and parametrize the coset as
\be
g(x) = e^{x^i P_i}e^{\Phi(x) S}e^{\Phi^i(x) S_i}\cdots e^{\Phi^{i_1\cdots i_N}(x)S_{i_1\cdots i_N}}~.
\label{arbitraryNcoset}
\ee
The corresponding Maurer--Cartan form is
\begin{align}
\nonumber
\omega =  g^{-1}\rd g=\rd x^i P_i &+(\rd\Phi+\Phi_i\rd x^i)S+(\rd\Phi^i+2\rd x^j\Phi_j^i)S_i+(\rd\Phi^{i_1i_2}+3\rd x^j\Phi_j^{i_1i_2})S_{i_1i_2}\\
&+\cdots+(\rd\Phi^{i_1\cdots i_{N-1}}+N\rd x^j\Phi_j^{i_1\cdots i_{N-1}})S_{i_1\cdots i_{N-1}}+\rd\Phi^{i_1\cdots i_N}S_{i_1\cdots i_N}.
\end{align}
The building blocks are the 1-forms
\be
\label{xNMaurer--Cartan1forms}
\omega_P^i = \rd x^i~,~~\omega_S = \rd\Phi+\Phi_i\rd x^i~~,\cdots,~~\omega_S^{i_1\cdots i_{K}}= \rd\Phi^{i_1\cdots i_{K+1}}+(K+1)\rd x^j\Phi_j^{i_1\cdots i_{K}},~~\cdots,~~\omega_S^{i_1\cdots i_N} =\rd\Phi^{i_1\cdots i_N}.
\ee

There are now $N$ inverse Higgs constraints,
\be
\rd\Phi^{i_1\cdots i_{K-1}}+K\rd x^j\Phi_j^{i_1\cdots i_{K-1}}=0~,\ \ \ K=1,\cdots,N
\ee
which implies
\be
\Phi_{i_1\cdots i_K} = -\frac{1}{K}\nabla_{(i_1}\Phi_{i_2\cdots i_K)}~,  \ \ \ K=1,\cdots,N
\ee
or, in terms of the scalar $\Phi$,
\be
\Phi_{i_1\cdots i_K} = \frac{(-1)^K}{K!}\nabla_{i_1}\cdots\nabla_{i_K}\phi~, \ \ \ K=1,\cdots,N.
\ee
After imposing these constraints, the basic building block is the form
\be
\omega_{S}^{i_1\cdots i_N}=\rd\Phi^{i_1\cdots i_N}= \frac{(-1)^N}{N!}\rd x^k\nabla_k\nabla^{i_1}\cdots\nabla^{i_{N}}\phi ,
\ee
which has $(N+1)$ derivatives and will thus be invariant under a shift with $N$ powers of $x$. As before, we are interested in constructing terms with fewer than $(N+1)$ derivatives per field, so we search for Wess--Zumino terms for this extended polynomial shift symmetry.

The exterior derivative acts on the basis forms as follows,
\begin{align}
\rd\omega_P^i &= 0,\\
\rd\omega_{S}^{i_1\cdots i_K} &= (K+1) \delta_{jk}\omega_{S}^{i_1\cdots i_K j}\wedge\omega_P^k ~,~~~~~~K=0,\cdots , N-1\\
\rd\omega_{S}^{i_1\cdots i_N} &= 0~.
\end{align}

\subsection{Tadpole and Quadratic Terms}

We do not need the elaborate higher-dimensional construction to find the invariant quadratic Lagrangians, so we will not write the explicit Wess--Zumino forms for this case.  The only possible non-trivial term with one field is the tadpole,
 \be
 {\cal L}_1\sim \phi~,
 \ee 
 and this is again a Wess--Zumino terrm, invariant up to a total derivative under the shifts \eqref{shiftsymmetryN}\footnote{The tadpole stems from the closed form \be
\omega_1 \sim\epsilon_{i_1\cdots i_D}\omega_S\wedge\omega_P^{i_1}\wedge\cdots\wedge\omega_P^{i_D}~.
\ee}.

With a single field the only possible quadratic Lagrangians are ${\cal L}_2\sim \phi\nabla^{2k}\phi$ for integer $k\geq0$.   Under an $N$-th order shift $\delta \phi\sim c_{i_1\cdot i_N}x^{i_1}\cdots x^{i_N}$ this term changes up to a total derivative by $\delta{\cal L}_2\sim c_{i_1\cdot i_N}\phi\square^k\left( x^{i_1}\cdots x^{i_N}\right)$ and this will vanish only if $2k>N$.  The cases $k\geq N+1$ are the cases where there are at least $N+1$ derivatives per field, {\it i.e.}, these are the coset constructible non-Wess--Zumino terms.  The Wess--Zumino terms are those with ${N\over 2}<k<N+1$,
\be {\cal L}_2\sim  \phi\nabla^{2k}\phi,\ \ \ ~~~~~~~{N\over 2}<k<N+1.\ee
For example, notice that the term
${\cal L}_2\sim \phi\nabla^4\phi~,$ which we constructed in Section \ref{N2quadtermssec} as a Wess--Zumino term for the $N=2$ symmetry $\delta\phi = c_{ij}x^ix^j$, is also a Wess--Zumino term for the $N=3$ cubic shift symmetry $\delta\phi = c_{ijk}x^ix^j x^k$.

\subsection{Interaction Terms}

For $N$ even, we can construct closed $(D+1)$-forms using only the highest form $\omega_S^{(N)}$,
\be 
\omega_n\sim\epsilon_{i_1\cdots i_D}\left[\omega_S^{(N)}\right]\wedge \left[\omega_S^{(N)}\wedge \omega_S^{(N)}\right]^{i_1i_2}\wedge\cdots\wedge \left[\omega_S^{(N)}\wedge \omega_S^{(N)}\right]^{i_{n-2}i_{n-1}}\wedge \omega_P^{i_n}\wedge\cdots\wedge\omega_P^{i_D},
\ee
where $\left[\omega_S^{(N)}\right]$ is the total trace of $\omega_S^{i_1\cdots i_N}$ and  $\left[\omega_S^{(N)}\wedge \omega_S^{(N)}\right]^{ij}\equiv \omega_{S\  k_1\cdots k_{N-1}}^{i}\wedge\omega_S^{j k_1\cdots k_{N-1}}$.
Going through the coset construction, this yields Lagrangians with odd $n$ powers of the field in dimension $D\geq n-1$ for $D$ even and $D\geq n$ for $D$ odd,
\be {\cal L}_n\sim  \epsilon_{i_1\cdots i_{n-1} k_{n}\cdots k_D}\epsilon^{j_1\cdots j_{n-1} k_{n}\cdots k_D}\nabla^{N}\phi \left[\nabla^{N+1}\phi\cdot \nabla^{N+1}\phi\right]^{i_1i_2}_{j_1j_2}\cdots\left[\nabla^{N+1}\phi\cdot \nabla^{N+1}\phi\right]^{i_{n-2}i_{n-1}}_{j_{n-2}j_{n-1}},\label{evenNlagint}\ee
where $\left[\nabla^{N+1}\phi\cdot \nabla^{N+1}\phi\right]^{i_1i_2}_{j_1j_2}\equiv \nabla^{i_1}\nabla_{j_1}\nabla_{ k_1}\cdots\nabla_{ k_{N-1}}\phi\nabla^{i_2}\nabla_{j_2}\nabla^{ k_1}\cdots\nabla^{ k_{N-1}}\phi$.

For $N$ odd, we can construct closed $(D+1)$-forms using the highest form $\omega_S^{i_1\cdots i_N}$ and the second highest form $\omega_S^{i_1\cdots i_{N-1}}$,
\be \omega_n\sim\epsilon_{i_1\cdots i_D}\left[\omega_S^{(N-1)}\right]\wedge \left[\omega_S^{(N)}\right]^{i_1}\wedge \cdots\wedge\left[\omega_S^{(N)}\right]^{i_{n-1}}\wedge \omega_P^{i_n}\wedge\cdots\wedge\omega_P^{i_D},\ee
where $\left[\omega_S^{(N-1)}\right]$ is the total trace of $\omega_S^{i_1\cdots i_{N-1}}$ and $\left[\omega_S^{(N)}\right]^i$ is the trace over all but one index of $\omega_S^{i_1\cdots i_{N}}$.  To see that this is closed, we use the fact that $\rd\left[\omega_S^{(N-1)}\right]\sim \left[\omega_S^{(N)}\right]_i\wedge \omega_P^i$, after which the resulting $(D+2)$-form, due to the structure of the epsilon tensor out front, always contains duplicates of some component of either $\omega_P^i$ or $\left[\omega_S^{(N)}\right]^i$, and hence vanishes by the anti-symmetry of the wedge product.  Going through the coset construction, this yields Lagrangians with all powers of the field $n$ up to $(D+1)$,
\be {\cal L}_n\sim  \epsilon_{i_1\cdots i_{n-1} k_{n}\cdots k_D}\epsilon^{j_1\cdots j_{n-1} k_{n}\cdots k_D}\nabla^{N-1}\phi  \nabla^{i_1}\nabla_{j_1}\nabla^{N-1}\phi\cdots \nabla^{i_{n-1}}\nabla_{j_{n-1}}\nabla^{N-1}\phi.\label{oddnlagint}\ee

A generic $(N+1)$ order Lagrangian can be expected to give $2(N+1)$-order equations of motion.  In these terms, \eqref{evenNlagint} and \eqref{oddnlagint}, there is a cancellation among the would-be highest derivative terms in the equations of motion, and we are left with $2N$ derivative equations.

One can also consider generalizations of the parity violating term \eqref{3dtotaldWZterm}, which we will not elaborate on here. 
Note that in the case of $N$ even, all the Wess--Zumino Lagrangians contain odd numbers of fields, but in the case of $N$ odd, there also exist terms with even numbers of fields, opening up the possibility of constructing theories with multiplets of fields transforming in the fundamental of some internal orthogonal group, and invariant under these extended polynomial shifts.

\section{Conserved Currents and Equations of Motion}

The symmetries \eqref{shiftsymmetryN} are global symmetries, and like all global symmetries they give rise to conserved currents via Noether's theorem.  In the case of shift symmetries, there are relations among the currents and the equations of motion, which can be used to generalize the observations of \cite{Nicolis:2010se}.

Given a Lagrangian ${\cal L}$ with an infinitesimal symmetry $\delta\phi$, we can find the associated Noether current $J^i$ by varying the action with a spacetime dependent coefficient in front of the symmetry $\phi\mapsto \phi+\alpha(x)\delta\phi$.  To lowest order in $\alpha(x)$,
\be \delta S=\int \rd^Dx {\delta{\cal L}\over \delta \phi}\,\alpha\,\delta\phi=-\int \rd^Dx \ \alpha \nabla_i J^i .\ee
In our case, the $N$-th order symmetry transformation is $\delta_{i_1\cdots i_N}\phi=x^{i_1}\cdots x^{i_N}$, and since $\alpha$ is arbitrary, we have the relation
\be  \nabla_i J_{(N)}^{i,i_1\cdots i_N}=-x^{i_1}\cdots x^{i_N}{\delta{\cal L}\over \delta \phi}.\label{djxeom}\ee
In particular, for the case $N=0$, {\it i.e.}, a theory with a standard shift symmetry, this implies that the equation of motion is the divergence of the Noether current associated with the shift symmetry: 
\be \nabla_i {J_{(0)}}^{i}=-{\delta{\cal L}\over \delta \phi}\label{zerodje}.\ee

As we will see now, in a theory with $N$-th order shift symmetry, the equations of motion can be expressed as the $(N+1)$-th derivative of a local $(N+1)$-index operator.  We define
\be 
{\cal S}_{(N)}^{i,i_1\cdots i_N}=\sum_{k=0}^N (-1)^k \left(\begin{array}{c}N \\r\end{array}\right) J_{(k)}^{i,(i_1\cdots i_k}x^{i_{k+1}}\cdots x^{i_N)}.\label{sdef}
\ee
The relative coefficients in this expression are such that after taking the $(N+1)$-th divergence and using \eqref{djxeom}, the terms with derivatives of the equations of motion all vanish and we are left with
\be \nabla_i\nabla_{i_1}\cdots\nabla_{i_N}{\cal S}_{(N)}^{i,i_1\cdots i_N}=-(-1)^N N! {\delta{\cal L}\over \delta \phi}.\label{dseom}\ee
In particular, ${\cal S}_{(0)}^{i}=J_{(0)}^{i}$ is the conserved shift symmetry and we have \eqref{zerodje}.
The equations of motion can thus be written the $(N+1)$-th derivative of a local $(N+1)$-index tensor, and hence integrated $(N+1)$ times.  
 
 In showing \eqref{dseom}, it helps to note that there is a recursion among the tensors \eqref{sdef},
 \be \nabla_i {\cal S}_{(N)}^{i,i_1\cdots i_N}=-N{\cal S}_{(N-1)}^{(i_1,i_2\cdots i_N)}.\ee
In particular, for $N=1$ we have that the Noether current for the shift symmetry is the derivative of a two index current.  \eqref{dseom} then follows by induction.

\section{Traceless Symmetry}

So far we have considered the symmetry
\be \phi\longmapsto \phi+c^{(0)}+c_i^{(1)}x^i+c_{ij}^{(2)}x^i x^j+\cdots+c_{i_1\cdots i_N}^{(N)}x^{i1}\cdots x^{i_N},\label{shiftsymmetry3}\ee  
where the $c$'s are constant, symmetric tensors, but are otherwise unrestricted.
For $N\geq 2$, however, we may consider smaller symmetry groups by putting some trace conditions on the $c$'s.
For example, for $N=2$, $c_{ij}^{(2)}$ can be broken into trace and trace-free components, and different invariant actions may be allowed depending on whether either or both components are present.  For the higher $N$'s we can separate out the single, double, triple, etc. traces,
\be c_{i_1\cdots i_N}^{(N)}\sim c_{i_1\cdots i_N}^{T}+\delta_{(i_1i_2}c_{i_3\cdots i_N)}^{T}+\delta_{(i_1i_2}\delta_{i_3i_4}c_{i_5\cdots i_N)}^{T}+\cdots,
\label{tracelesssymm}
\ee
where the ${c^T}$'s are all traceless.  We can consider various possibilities depending on which of these components are present.

In the next subsection we will consider first the specific case of a traceless $N=2$ symmetry, and then the more general case of a completely traceless $N$-th order symmetry, showing that the standard two-derivative kinetic term is a Wess--Zumino term with respect to these symmetries.  We will not attempt here a general study of all the possible interaction terms.  

\subsection{Traceless $N=2$}
\label{tracelessx2}

Here we consider the case $N=2$ where the quadratic shift symmetry is traceless, $\delta^{ij}c^{(2)}_{ij}=0$.
In this case, the infinitesimal symmetries are
\be
\delta_C\phi = 1~,~~~~~~~~~~~~~~~~~~~\delta_{B^i}\phi = x^i~,~~~~~~~~~~~~~~~~~~~\delta_{\tilde S^{ij}}\phi = x^i x^j - \frac{1}{D}x^2\delta^{ij}~.
\ee
The commutation relations are now
\begin{align}
\nonumber
\label{x2algebrat}
\left [P_i,B_j\right ] &= \delta_{ij }C~,  ~~~~~~~~~~~~~~~~~~~~~~~~~~~~~~ \left[P_{i },\tilde S_{jk}\right] = \delta_{ij}B_k+\delta_{ik}B_j-\frac{2}{D}B_i\delta_{jk}~, \\
\left[J_{ij},B_{k }\right]& = \delta_{ik }B_{j }-\delta_{jk}B_{i}~,~~~~~~~~~~~~~~~~~~ \left[J_{ij},\tilde S_{kl},\right] =\delta_{ ik}\tilde S_{jl}- \delta_{ jk }\tilde S_{ il }+\delta_{  il}\tilde S_{ k j}-\delta_{  jl}\tilde S_{ k i}~, \\\nonumber
\left [J_{ij },P_{k }\right ] &=\delta_{ik}P_{j } - \delta_{jk }P_{i }~,~~~~~~~~~~~~~~~~~~~\left [J_{ij },J_{kl }\right ] = \delta_{ik}J_{jl }-\delta_{jk }J_{il }+\delta_{jl}J_{ik }-\delta_{il}J_{jk }~.
\end{align}
The broken symmetries are those corresponding to the generators $C, B_i, \tilde S_{ij}$, with corresponding field $\phi(x),\xi^i(x)$ and traceless $\tilde \Phi_{ij}(x)$, and the coset is parameterized by
\be
g(x) = e^{x^i P_j}e^{\phi(x) C}e^{\xi^i(x) B_i}e^{\tilde \Phi^{ij}(x) \tilde S_{ij}}~.
\ee
The Maurer--Cartan forms look the same as in Section \eqref{N2secsta} because the modification to the algebra projects out,
\begin{align}
\omega_P^i &= \rd x^i,\\
\omega_C &= \rd\phi+\xi_i\rd x^i,\\
\omega_{\tilde S}^{ij} &= \rd \tilde \Phi^{ij},\\
\omega_B^i &= \rd\xi^i+2\rd x^j \tilde \Phi_j^i~.
\end{align}
However, there are differences: $\omega_{\tilde S}^{ij}$ is now traceless, and the form that one of the inverse Higgs constraints takes is now slightly different.  The relation coming from the commutator, $\left [P_i,B_j\right ] = \delta_{ij}C$, is the same as before and takes the form
\be
\xi_i= -\nabla_i\phi~,
\label{tracelessinversehiggs1}
\ee
but the commutator $\left[P_{i },\tilde S_{jk}\right] = \delta_{ij}B_k+\delta_{ik}B_j-\frac{2}{D}B_i\delta_{jk}~$
now tells us that we should set the traceless part of the form $\omega_B^i$ equal to zero. Writing
\be
\omega_B^i = \rd x^j \left(\nabla_j\xi^i +2 \tilde \Phi_j^i\right)~,
\ee
this translates to the requirement that
\be
\nabla_i\xi_j +\nabla_j\xi_i+4 \tilde \Phi_{ij} -\frac{2}{D}\nabla_k\xi^k \delta_{ij} = 0,
\ee
which leads to the following relation for $\tilde \Phi_{ij}$,
\be
\tilde \Phi_{ij} = \frac{1}{2D}\nabla_k\xi^k \delta_{ij} - \frac{1}{2}\nabla_{(i}\xi_{j)}  = \frac{1}{2}\nabla_i\nabla_j\phi-\frac{1}{2D} \nabla^2\phi \delta_{ij}~,
\label{tracelessinversehiggs2}
\ee
where in the last equality we have substituted in the relation~\eqref{tracelessinversehiggs1}. Notice that we have only set the traceless part of $\omega_B$ equal to zero, therefore, when we substitute~\eqref{tracelessinversehiggs1} back into $\omega_B$, we will isolate its trace. Similarly, we can substitute the inverse Higgs constraints into $\omega_{\tilde S}$ to obtain the invariant building blocks
\begin{align}
\omega_B^i &= -\frac{1}{D}\rd x^i\nabla^2\phi,\\
\omega_{\tilde S}^{ij} &= \frac{1}{2}\rd x^k\left(\nabla_k\nabla^i\nabla^j\phi - \frac{1}{D}\nabla_k\nabla^2\phi \delta^{ij}\right)~.
\end{align}
Notice that now in addition to the expected three derivative building block $\nabla\nabla\nabla\phi$, we also can use the laplacian $\nabla^2\phi$ to build invariant Lagrangians.

The fact that $\nabla^2\phi$ is one of the building blocks for the traceless symmetry means that the Wess--Zumino terms we constructed for the quadratic shift symmetry are now all (with the exception of the parity violating term \eqref{3dtotaldWZterm}){ strictly} invariant terms, constructible from the coset.  Thus they are not Wess--Zumino with respect to the traceless version of the symmetry.

However, we now have new Wess--Zumino terms available.  Consider
\be\label{tracelessform2}
\tilde \omega_2 =  \epsilon_{i_1\cdots i_{D}} \omega_C\wedge\omega_B^{i_1}\wedge\omega_P^{i_2}\wedge\cdots\wedge\omega_P^{i_D}~.
\ee
This form is now closed,
\be
\rd\tilde\omega_2 \sim \epsilon_{i_1\cdots i_{D}} \omega_C\wedge\omega_{S\ k}^{i_1 }\wedge\omega_P^k\wedge\omega_P^{i_2}\wedge\cdots\wedge\omega_P^{i_D} \sim \omega_C\wedge[\omega_S] \rd ^Dx= 0~,
\ee
where we have used that $\omega_S$ is traceless.
This form is not the exterior derivative of a left-invariant form, so it will have a corresponding Wess--Zumino term.  Writing $\tilde\omega_2$ out,
\be
\tilde\omega_2\sim \epsilon_{i_1\cdots i_{D}}(\rd\phi+\xi_k\rd x^k)\wedge(\rd\xi^{i_1}+2\rd x^\ell \tilde \Phi_\ell^{i_1})\wedge\rd x^{i_2}\wedge \cdots\wedge\rd x^{i_D}~.
\ee
The term involving $\tilde \Phi$ is $\propto [\tilde \Phi]$, which is zero, so we can write this as $\tilde\omega_2 = \rd\tilde\beta_2$ with
\be
\tilde\beta_2 \sim  \epsilon_{i_1\cdots i_{D}}\phi~ \rd\xi^{i_1}\wedge\rd x^{i_2}\wedge \cdots\wedge\rd x^{i_D} -\frac{1}{2D}\epsilon_{i_1\cdots i_{D}} \xi^2\rd x^{i_1}\wedge\rd x^{i_2}\wedge \cdots\wedge\rd x^{i_D} ~,
\ee
which upon pulling back, integrating, substituting the inverse Higgs constraint \eqref{tracelessinversehiggs1}, and integrating by parts, leads to a Lagrangian which is nothing but the standard two-derivative kinetic term,
\be\label{kineticterm1}
\tilde {\cal L}_2 \sim  \phi\nabla^2\phi~.
\ee
It is straightforward to check that this term is indeed invariant up to a total derivative under a traceless quadratic shift.

\subsection{Traceless Symmetry for Arbitrary $N$}

We can extend the traceless symmetry to arbitrary $N$.  We consider here only the fully traceless component of the symmetry~\eqref{shiftsymmetry3}, corresponding to
\be
\delta_{\tilde S^{i_1\cdots i_K}}\phi = x^{(i_1}x^{i_2}\cdots x^{i_K)_T}~,~~~~ K=0,\cdots,N \label{tracelessN}
\ee
where $(~)_T$ indicates the symmetric traceless part of the enclosed indices. This corresponds to taking
$c_{i_1\cdots i_N}^{(K)} \sim c_{i_1\cdots i_K}^{T}~$
in~\eqref{tracelesssymm}, for each of the symmetries. The commutation relations for this traceless symmetry are
\begin{align}
[P_j, \tilde S_{i_1\cdots i_K}] &= \sum_{\ell=1}^K\delta_{j ( i_\ell}\tilde S_{i_1\cdots \hat i_\ell\cdots i_K)_T},\ \ \ ~~~~~~~K=0,\cdots,N \nn\\
[J_{jk}, \tilde S_{i_1\cdots i_K}] &= \sum_{\ell=1}^K\left(\delta_{j i_\ell}\tilde S_{k i_1\cdots \hat i_\ell\cdots i_K} - \delta_{k i_\ell}\tilde S_{j i_1\cdots \hat i_\ell\cdots i_N}\right),\ \ \ ~~~~K=0,\cdots,N \nn\\
\left [J_{ij },P_{k }\right ] &=\delta_{ik}P_{j } - \delta_{jk }P_{i },~~~~~~~~~~~~\left [J_{ij },J_{kl }\right ] = \delta_{ik}J_{jl }-\delta_{jk }J_{il }+\delta_{jl}J_{ik }-\delta_{il}J_{jk }~.
\end{align}
The coset element takes the same form as~\eqref{arbitraryNcoset}, and the Maurer--Cartan forms look the same as \eqref{xNMaurer--Cartan1forms} (as in the quadratic case, the difference in the algebras projects out)
\begin{align}
\nonumber
\omega =\rd x^i P_i &+(\rd\tilde\Phi+\tilde\Phi_i\rd x^i)\tilde S+(\rd\tilde\Phi^i+2\rd x^j\tilde\Phi_j^i)\tilde S_i+(\rd\tilde\Phi^{i_1i_2}+3\rd x^j\tilde\Phi_j^{i_1i_2})\tilde S_{i_1i_2}\\
&+\cdots+(\rd\tilde\Phi^{i_1\cdots i_{N-1}}+N\rd x^j\tilde\Phi_j^{i_1\cdots i_{N-1}})\tilde S_{i_1\cdots i_{N-1}}+\rd\tilde\Phi^{i_1\cdots i_N}\tilde S_{i_1\cdots i_N}~,
\end{align}
\be
\label{xNMaurer--Cartan1formsaN}
\omega_P^i = \rd x^i~,~~\omega_{\tilde S} = \rd\tilde\Phi+\tilde\Phi_i\rd x^i~~,\cdots,~~\omega_{\tilde S}^{i_1\cdots i_{K}}= \rd\tilde\Phi^{i_1\cdots i_{K+1}}+(K+1)\rd x^j\tilde\Phi_j^{i_1\cdots i_{K}},~~\cdots,~~\omega_{\tilde S}^{i_1\cdots i_N} =\rd\tilde\Phi^{i_1\cdots i_N},
\ee
where each of the $\tilde\Phi_{i_1\cdots i_K}$ are traceless.
The inverse Higgs constraints now tell us that we can set the traceless parts of the Maurer--Cartan forms to zero
\be
\partial_{(j}\tilde\Phi_{i_1\cdots i_{K-1})_T}+K\tilde\Phi_{(ji_1\cdots i_{K-1})_T}=0~,\ \  ~~~~~ K=1,\cdots,N.
\ee
Solving these constraints implies that
\be
\tilde\Phi_{i_1\cdots i_K} = -\frac{1}{K}\nabla_{(i_1}\tilde\Phi_{i_2\cdots i_K)_T}~,  \ \ \ K=1,\cdots,N~,
\ee
and substituting them back yields two invariant building blocks:
\begin{align}
\omega_{\tilde S}^i &= -\frac{1}{D}\rd x^i\nabla^2\phi~,\\
\omega_{\tilde S}^{i_1\cdots i_N} &= \rd\tilde\Phi^{i_1\cdots i_N}= \frac{(-1)^N}{N!}\rd x_k\nabla^{(k}\nabla^{i_1}\cdots\nabla^{i_{N})_T}\phi.
\end{align}

The kinetic term \eqref{kineticterm1} is invariant up to a total derivative under a fully-traceless polynomial shift for any $N$,
$\phi=\sum_{N=0}^\infty c^T_{i_1\cdots i_N}x^{i_1}\cdots x^{i_N}~.$
(In fact, the conserved charges of this symmetry are nothing but the standard multipole coefficients, see Appendix~\ref{multipoleapp}.)
The term
\be
\tilde \omega_2 =  \epsilon_{i_1\cdots i_{D}} \omega_C\wedge\omega_B^{i_1}\wedge\omega_P^{i_2}\wedge\cdots\wedge\omega_P^{i_D}~,
\ee
continues to be a Wess--Zumino term for the arbitrary $N$ traceless symmetry, and leads to the kinetic term \eqref{kineticterm1},
\be
\tilde{\cal L}_2 \sim \phi\nabla^2\phi~,
\ee
which is invariant up to a total derivative under~\eqref{tracelessN}.

\section{Conclusions}
We have studied interacting scalar field Lagrangians which are invariant under extended polynomial shift symmetries, with emphasis on the terms which shift only by a total derivative under these symmetries.  These are the Wess--Zumino terms, which can be thought of as the natural generalization of the tadpole term for constant shifts and the galileon terms for shifts linear in the coordinates.  They have fewer derivatives per field and lower order equations of motion than the strictly invariant terms.

Due to their Wess--Zumino nature, we expect these terms to inherit some of the attractive properties of the galileon terms.   In the relativistic case, an extension of the derivative-counting argument in \cite{Hinterbichler:2010xn} shows that these terms will not undergo quantum corrections to any order in perturbation theory.   Similarly, we expect that they will exhibit Vainshtein-like screening around massive sources \cite{Vainshtein:1972sx} (for introductions to screening, see~\cite{Jain:2010ka,Babichev:2013usa,Khoury:2013tda}). However, the study of the effective theory of which these terms are a part is a somewhat intricate task, especially for the non-relativistic case which seems necessary to avoid ghostly degrees of freedom.   In addition, it would be interesting to couple these terms to gravity and understand what cosmological consequences they may have.  In the non-relativistic case, it is likely that these theories will have to be coupled to something along the lines of Ho\v{r}ava--Lifshitz gravity~\cite{Horava:2009uw}. We anticipate that these theories may prove useful also in the condensed matter context outlined in \cite{Griffin:2013dfa}, where they may describe systems in the neighborhood of critical points.

\vspace{.2cm}
\noindent
{\large {\bf Acknowledgements:}} We would like to thank Garrett Goon and Justin Khoury for helpful discussions. AJ would like to thank the Perimeter Institute for hospitality both when this work was initiated and when it was completed. Research at Perimeter Institute is supported by the Government of Canada through Industry Canada and by the Province of Ontario through the Ministry of Economic Development and Innovation. This work was made possible in part through the support of a grant from the John Templeton Foundation. The opinions expressed in this publication are those of the authors and do not necessarily reflect the views of the John Templeton Foundation (KH). This work was supported in part by the Kavli Institute for Cosmological Physics at the University of Chicago through grant NSF PHY-1125897, an endowment from the Kavli Foundation and its founder Fred Kavli, and by the Robert R. McCormick Postdoctoral Fellowship (AJ).

\vspace{.7cm}
\noindent
{\Large\bf Appendix}
\appendix
\section{Multipole Moments as Noether Charges for Extended Traceless Shift Symmetries}
\label{multipoleapp}

Here we remark that the extended shift symmetries considered are not really so alien, and that they appear (with associated physical consequences) already in a quite familiar situation: a free scalar field coupled to a source in three dimensions. Consider the Lagrangian
\be {\cal L}={1\over 2}(\nabla\phi)^2+\rho\phi,\ee
in $D=3$ Euclidean dimensions, for some scalar field $\phi(x)$ and external source $\rho(x)$.  This yields the Poisson equation 
\be \nabla^2\phi=\rho.\ee
This Lagrangian has an extended traceless shift symmetry for all $N$, 
\be \delta \phi=\sum_{N=0}^\infty c_{i_1\cdots i_N}x^{i_1}\cdots x^{i_N},\ee
where $c_{i_1\cdots i_N}$ is a symmetric traceless matrix of coefficients.   This is a symmetry even in the presence of $\rho$ (since $\rho\phi$ shifts to a fixed function, and any fixed function is a total derivative).  Like any symmetry, there is a conserved Noether current.  Outside the source, this Noether current is
\be j=\nabla\phi\, \delta\phi-\phi\nabla\delta\phi.\ee

The conserved charge can be evaluated on any closed 2-surface with normal $\hat n$, and is independent of the choice of surface,
\be Q=\oint j\cdot \hat n.\label{ccharge}\ee
In particular, we may evaluate it on a sphere of radius $r\equiv \sqrt{x^ix_i}$ which is large enough so that all of its points are outside the source.  For this we may use the standard multipole expansion of the field in terms of the source,
\be \phi(x)=-{1\over 4\pi  }\sum_{N=0}^\infty {1\over r^{2N+1}}  x^{i_1} \cdots  x^{i_N} Q_{i_1\cdots i_N},\label{multipoleex}\ee
where the $Q_{i_1\cdots i_N}$ are the standard multipole moments of the charge distribution defined by
\be Q_{i_1\cdots i_N}={(2N)!\over 2^N(N!)^2}\int \rd^3\xb' \ (x'^{i_1} \cdots  x'^{i_N})_{\rm T} \rho(\xb'),\ee
where $(\ )_{\rm T}$ refers to the traceless part.

Evaluating the conserved charge \eqref{ccharge} using \eqref{multipoleex}, we see that the multipole moments are nothing but the conserved charges associated to the higher shift symmetries:\footnote{The multipole moments can also be thought of as the Noether charges associated to conformal Killing tensors~\cite{Geroch:1970cc}.}
\be Q=\int r^2 \rd\Omega\ \partial_r\phi\delta\phi-\phi\partial_r \delta\phi=-{2^N(N!)^2\over (2N)! }Q_{i_1\cdots i_N}c^{i_1\cdots i_N}.\ee

\renewcommand{\em}{}
\bibliographystyle{utphys}
\addcontentsline{toc}{section}{References}
\bibliography{gengalarxiv}

\end{document}